\documentclass[twocolumn,nofootinbib,superscriptaddress,10pt,aps,prd]{revtex4-1}

\usepackage{mathtools}
\usepackage{amsmath}
\usepackage{amssymb}
\usepackage[utf8]{inputenc}
\usepackage[T1]{fontenc}
\usepackage[english]{babel}
\usepackage{graphicx}
\usepackage{epsfig}
\usepackage{epstopdf}
\usepackage{dcolumn}
\usepackage{bm}
\usepackage{color}
\usepackage{tabularx}
\usepackage{multirow}
\usepackage{hyperref}
\hypersetup{colorlinks=true,
            linkcolor=blue,
            citecolor=blue,
            urlcolor=blue}
\usepackage{slashed}
\usepackage{physics}

\begin{document}

\title{Parton distribution and fragmentation functions with massive gluons}

\author{Gustavo B. Bopsin}
\email{gustavo.bopsin@unesp.br}
\affiliation{Instituto de F\'{\i}sica Te\'orica, Universidade Estadual Paulista, Rua Dr.~Bento Teobaldo Ferraz 271 - Bloco II, 01140-070 S\~ao Paulo, S\~ao Paulo, Brazil}

\author{Bruno El-Bennich}
\email{bennich@unifesp.br}
\affiliation{Instituto de F\'{\i}sica Te\'orica, Universidade Estadual Paulista, Rua Dr.~Bento Teobaldo Ferraz 271 - Bloco II, 01140-070 S\~ao Paulo, S\~ao Paulo, Brazil}
\affiliation{Departamento de F\'isica, Universidade Federal de S\~ao Paulo, Rua S\~ao Nicolau 210, 09913-030 Diadema, S\~ao Paulo, Brazil}

\author{Gast\~ao Krein}
\email{gastao.krein@unesp.br}
\affiliation{Instituto de F\'{\i}sica Te\'orica, Universidade Estadual Paulista, Rua Dr.~Bento Teobaldo Ferraz 271 - Bloco II, 01140-070 S\~ao Paulo, S\~ao Paulo, Brazil}

\author{Fernando E. Serna}
\email{fernando.serna@unisucrevirtual.edu.co}
\affiliation{Departamento de F\'isica, Universidad de Sucre, Carrera 28 No.~5-267, Barrio Puerta Roja, Sincelejo 700001, Colombia}

\author{Roberto C. da Silveira}
\email{rc.silveira@unesp.br}
\affiliation{Instituto de F\'{\i}sica Te\'orica, Universidade Estadual Paulista, Rua Dr.~Bento Teobaldo Ferraz 271 - Bloco II, 01140-070 S\~ao Paulo, S\~ao Paulo, Brazil}


\begin{abstract}
The correct description of the hadron’s structure requires understanding how quarks and gluons form the observable hadrons and how they are distributed within them. Two key nonperturbative quantities encapsulate this information: parton distribution functions (PDFs) and fragmentation functions (FFs). The former define a probabilistic light-front momentum distribution of partons within a hadron, whereas the latter describe the hadronization process of high-energy partons. Computing these functions analytically poses significant challenges, as it demands models that accurately incorporate the nonperturbative infrared dynamics of Quantum Chromodynamics (QCD). In this work, we compute the pion PDF and its elementary and full FFs using the Curci-Ferrari (CF) model. This model enables the exploration of nonperturbative QCD effects by introducing a gluon-mass scale within the Landau gauge QCD Lagrangian. The two-point quark and gluon correlation functions derived from the CF model agree well with lattice QCD results and reproduce the pion decay constant in the chiral limit, consistent with chiral perturbation theory. The resulting pion PDF and FFs computed with the CF quark propagator and pion Bethe-Salpeter amplitude are in good qualitative and quantitative agreement with those obtained using the Qin-Chang model, a benchmark approach to nonperturbative QCD. These findings support the broader applicability of the Curci-Ferrari model in hadron phenomenology. 
\end{abstract}

\maketitle


\section{Introduction}

Parton distribution functions (PDFs) and fragmentation functions (FFs) play a crucial role in describing the internal structure of hadrons and their formation in high-energy processes. 
While the former characterize the momentum distributions of partons inside a hadron, the latter describe how a high-energy parton hadronizes into observable particles. Both objects arise in the factorization of hard scattering cross sections in Quantum Chromodynamics (QCD), allowing for the separation of perturbative short-distance and nonperturbative long-distance 
effects~\cite{Collins:1998be}. Thus, these distributions are inherently of nonperturbative nature and the past years have seen considerable efforts to obtain them indirectly in Euclidean space with lattice-QCD simulations~\cite{Oehm:2018jvm,Holligan:2024umc,Alexandrou:2020gxs,Gao:2020ito} and in the framework of the Dyson-Schwinger equation (DSE)~\cite{Nguyen:2011jy,Bednar:2018mtf,daSilveira:2024ddq}, with the Nambu--Jona-Lasinio (NJL)~\cite{Bentz:1999gx,Ito:2009zc,Courtoy:2019cxq} and the nonlocal chiral quark models~\cite{Nam:2011hg}. 

PDFs are defined in the operator product expansion of the hadronic tensor $W_{\mu\nu}$ that enters the cut diagram of the forward Compton amplitude~\cite{Jaffe:1983hp} 
applying the optical theorem. In particular, the cut enforces an on-shell condition for the quarks, thereby ensuring that only physical partons contribute to the process and allowing for a probabilistic interpretation of the partons. This diagrammatic approach has taken advantage of using quark propagators and Bethe-Salpeter wave functions in both the NJL model~\cite{Bentz:1999gx,Ito:2009zc,Courtoy:2019cxq} and the functional DSE approach~\cite{Nguyen:2011jy,Bednar:2018mtf,daSilveira:2024ddq}. Moreover, the quark-to-pion FF and the PDF of a valence quark in a pion obey a Drell-Levy-Yan (DLY) relation, as charge conjugation and crossing symmetry relate both functions in different kinematic regions~\cite{Drell:1969jm,Drell:1969wd,Drell:1969wb}. A notable feature of Ref.~\cite{daSilveira:2024ddq} is that the DSE and NJL predictions for the elementary quark-to-pion FF are rather alike despite the point-like nature of the pion's wave function in the latter model. More precisely, the FF obtained in the DSE framework is about 10--12\% enhanced for  momentum fractions $z \lesssim 0.8$, when compared to an earlier NJL model calculation~\cite{Ito:2009zc}, but otherwise in good qualitative agreement.

A common feature of the NJL model and the DSE applied to QCD is \emph{dynamical chiral symmetry breaking\/}, that is, the quark gap equation possesses a nonperturbative solution even in the chiral limit. The strength of the DSE kernel comes from both the quark-gluon vertex and the fully dressed gluon propagator, which obeys its own gap equation. Massive solutions to the latter exist and have been the subject of intense debates~\cite{Alkofer:2003jj,Cucchieri:2009zt,Bogolubsky:2009dc,Boucaud:2008ky,Duarte:2016iko,Dudal:2018cli, 
Duarte:2017wte,Aguilar:2004sw,Aguilar:2008xm}. In particular, the \emph{decoupling\/} solution of the gluon DSE and the associated effective gluon mass have inspired calculations with a higher-derivative Podolsky Lagrangian~\cite{El-Bennich:2020aiq} and in a particular version~\cite{Tissier:2011ey,Pelaez:2013cpa,Pelaez:2014mxa,Pelaez:2017bhh,Gracey:2019xom,Reinosa:2020skx,Barrios:2021cks} of the Curci-Ferrari (CF) model~\cite{Curci:1976bt}. In the latter model, the standard Faddeev-Popov QCD Lagrangian is augmented by a gluon-mass term. This term clearly violates the Becchi-Rouet-Stora-Tyutin (BRST) symmetry of the Faddeev-Popov Lagrangian, though it is still symmetric under a modified BRST symmetry responsible for its renormalizability. One may also think of this procedure as a minimal extension to take into account the effects of Gribov copies~\cite{Reinosa:2020skx}. 

The CF model has been used to explore the infrared behavior of Green functions in QCD~\cite{Tissier:2011ey,Pelaez:2013cpa,Pelaez:2014mxa,Pelaez:2017bhh,Gracey:2019xom,Reinosa:2020skx,Barrios:2021cks,Pelaez:2015tba,Barrios:2020ubx,Figueroa:2021sjm,Barrios:2022hzr,Barrios:2024ixj}, and at two-loop order the dressed ghost and gluon propagators in Landau gauge are in rather good qualitative agreement with the propagators obtained in lattice QCD~\cite{Cucchieri:2009zt,Bogolubsky:2009dc,Boucaud:2008ky,Duarte:2016iko,Duarte:2017wte,Dudal:2018cli} and with the gluon DSE~\cite{Aguilar:2004sw,Aguilar:2008xm}. However, as dynamical chiral symmetry breaking in the chiral limit is truly a nonperturbative phenomenon, even next-to-leading order perturbation theory only reproduces predicted quark-mass functions  with a large gluon mass and an unphysical current-quark mass~\cite{Barrios:2021cks}. An alternative approach combines an expansion in the Yang-Mills coupling and in the inverse number of colors $N_c$, without expanding in the quark-gluon coupling~\cite{Pelaez:2017bhh,Pelaez:2020ups,Pelaez:2022rwx}, also known as rainbow-improved loop expansion. At leading order in this expansion, a quark-antiquark bound state corresponds to the resummation of a ladder diagram of massive gluon exchanges. Fitting the CF model parameters to the lattice QCD predictions of the gluon, ghost, and quark dressing functions, a mass function is predicted that yields a reasonable Euclidean constituent quark mass and the perturbative tail of the mass function is qualitatively in agreement with the lattice data. 

Our goal in this work is to apply and evaluate the capability of the CF model to describe physical observables beyond two-point Green functions. To this end, we employ dressed quark propagators and Bethe-Salpeter amplitudes~(BSAs), 
derived in the CF model in the chiral limit~\cite{Pelaez:2022rwx}, to compute the PDF and FF of a pion with the aforementioned cut diagram. Finally, we compare both distribution functions with those of a DSE calculation~\cite{daSilveira:2024ddq} to assess whether the CF model reproduces their qualitative and quantitative behavior as a function of the quark's momentum fraction. 

This paper is organized as follows. Section~\ref{sec:pdf_frag} introduces the definitions of the PDF and FF, and discusses the quark-jet model. Section~\ref{sec:framework} briefly summarizes the DSE formalism with both the QC and CF gluon models. Our numerical results are discussed in  Section~\ref{sec:numericals} and conclusions are given in Section~\ref{sec:conclusions}.


\section{Distributions and Fragmentation Functions} 
\label{sec:pdf_frag}

The quest of understanding the internal structure of hadrons has led to the definitions of nonperturbative functions that encode how quarks and gluons are distributed inside a hadron and how they hadronize into observable particles~\cite{Aguilar:2024otb}. Amongst others, two fundamental one-dimensional quantities fulfill this task: PDFs and FFs. The former describes the light-front momentum distribution of partons within a hadron, while the latter describes their hadronization process. These functions are central to factorization theorems and the interpretation of high-energy scattering experiments.


\subsection{Parton Distribution Function}
\label{PDFsubsec}

\begin{figure}[t!] 
\centering
  \includegraphics[width=0.35\textwidth]{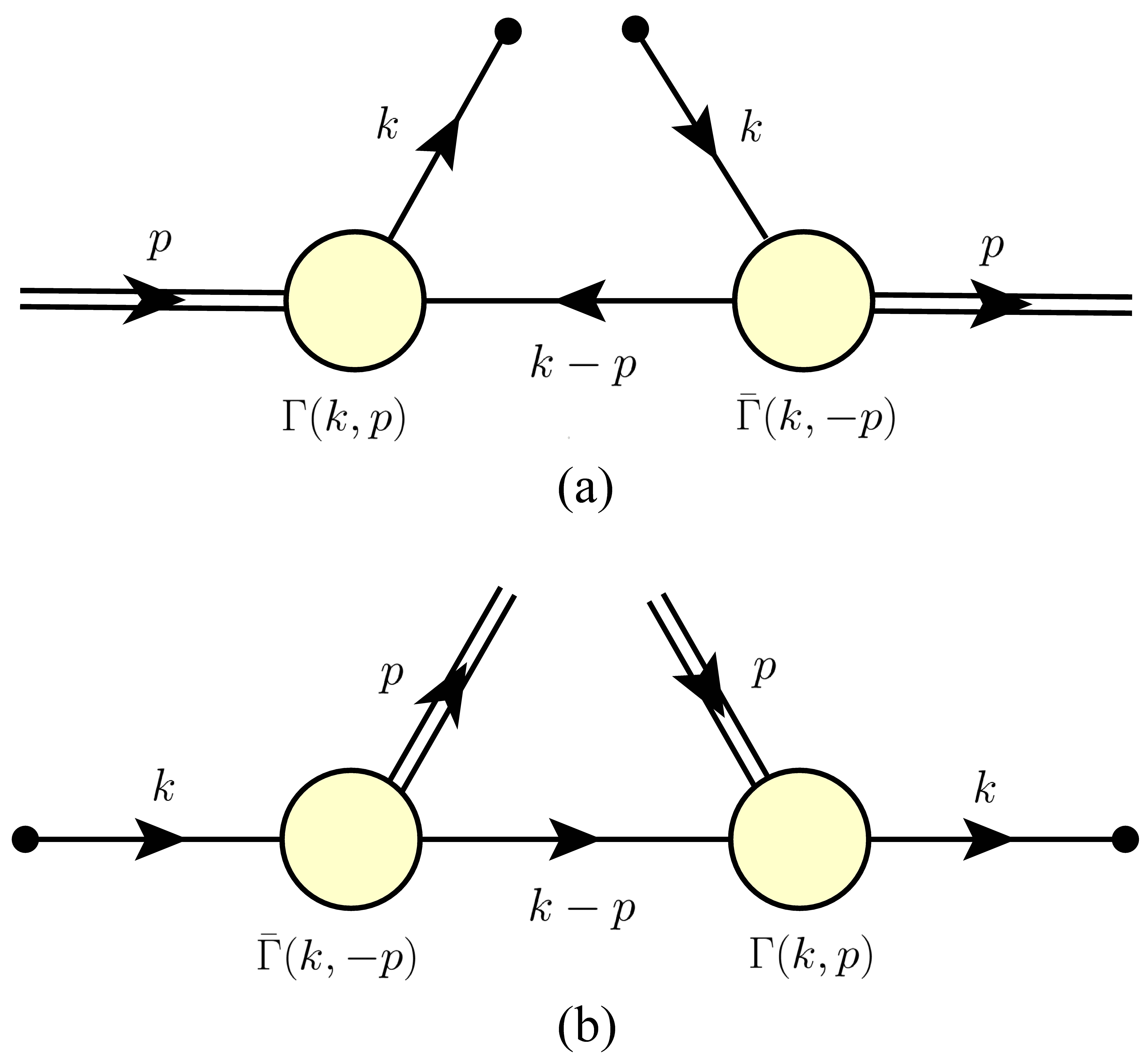} 
\caption{Cut-diagram representation of the parton distribution functon \protect\hypertarget{fig:diagram_a}{(a)} and elementary fragmentation function \protect\hypertarget{fig:diagram_b}{(b)}. The shaded circle with outgoing/incoming double-solid lines denote the pion in the fragmentation process, while solid lines are quark propagators and solid dots represent the $\gamma^+$ between the two quarks with momentum $k$.  }
\label{diagram} 
\end{figure}

The PDF of a hadron, $h$, is defined through the hadronic matrix element of a quark-field bilinear operator, 
\begin{eqnarray}
   \hspace*{-3mm}
    f_q^h(x) = \int \frac{d\omega^-}{4\pi} e^{i x p_- \omega^-} \! \bra{p(h)} \bar{\psi}(0) \gamma^+ \psi(\omega^-) \ket{p(h)} \!,
   \label{pdf_field_def}   
\end{eqnarray}
where $x =k^+/p^+$ is defined as the light-front momentum fraction of a quark with momentum $k$ in a hadron with total momentum $p$.
We compute this distribution function with the relevant cut diagram~\cite{Ito:2009zc,Bentz:1999gx,Jaffe:1983hp} and use the conventions of Ref.~\cite{Eichmann:2021vnj}
for the  light-cone variables in Euclidean space. The Feynman rules of the cut diagram in Euclidean metric yield~\cite{Ito:2009zc},
\begin{align}
   f^\pi_q(x )= &\, N_c C_q^\pi \int_k \delta \qty(k^+-xp^+)  \, \Tr_\mathrm{D}  \big [ S(k)\gamma^+S(k)  \hspace*{3mm}  \nonumber \\
  &  \times\,  \bar{\Gamma}_\pi \qty(-k + \tfrac{p}{2},-p)  S_\mathrm{os} (l) \Gamma_\pi \qty(k-\tfrac{p}{2},p) \big  ]  , 
  \label{pdf_func}   
\end{align}
where $\int_k = \int d^4k/(2\pi)^4$,  $l = k - p$, $N_c = 3$, $S (k)$ is a dressed quark propagator, $S_{\mathrm{os}}(l)$ is the on-shell quark propagator, $\Gamma_\pi\qty(k,p)$ is the BSA
of the pion and $C_q^\pi = \frac{1}{2} (1 + \tau_q \tau_\pi)$ is a product of isospin factors using the basis $(\tau_u, \tau_d) = (1, -1)$ and $(\pi^+, \pi^0, \pi^-) = (1, 0, -1)$.  
In the on-shell propagator,
\begin{equation*}
   S_\mathrm{os} (l) = 2\pi i  \delta \big(l^2+M^2_E\big) \! \left [  -i \gamma \cdot \ell  + M (\ell^2 ) \right ] ,
\end{equation*}
we define a Euclidean constituent mass with $M^2(k^2) = k^2$~\cite{Serna:2018dwk}. The relevant cut diagram for the PDF is illustrated in Fig.~\hyperlink{fig:diagram_a}{\ref{diagram}(a)} and the PDF is normalized as:
\begin{equation}
   \int_0^1 \hat{f}_u^\pi (x) \,dx = 1 \;  .
\label{pdf_sum_rule}   
\end{equation}

%

\subsection{Elementary Fragmentation Functions}

A simple relation exists between the PDF for unphysical $x>1$ and the elementary FF for physical $z\leq1$,  based on charge conjugation and crossing symmetry, namely the called Drell-Levy-Yan relation~\cite{Drell:1969jm,Drell:1969wd,Drell:1969wb}:
\begin{eqnarray}
\label{dly_relation1}
d_q^h(z)=(-1)^{2(s_q+s_h)+1}\frac{z}{d_q}f_q^h\qty(x=\frac{1}{z}) ,
\end{eqnarray}
where $s_q$ and $s_h$ are the spins of the quark and hadron, respectively, and $d_q$ demotes the spin-color degeneracy of the quark. Applied to the pion FF we have
$d_q=6$,  $s_q\equiv s_u=1/2$ and $s_h=s_\pi=0$, so that Eq.~\eqref{dly_relation1} becomes,
\begin{eqnarray}
\label{dly_relation2}
  d_q^\pi(z)  =  \dfrac{z}{6}f_q^\pi\qty(x=\frac{1}{z})\, ,
\end{eqnarray}
which yields:
\begin{align} 
   d_{q}^{\pi} (z) = & \, \frac{N_c   C_q^\pi z}{6}\! \int_k \delta\Big (k^{+} - \tfrac{p^+}{z} \Big ) \operatorname{Tr}_\text{D}  \Big [  S(k)\gamma^{+}S(k) 
     \hspace*{3mm}   \nonumber \\
   & \times \bar{\Gamma}_{\pi} \big (-k+\tfrac{p}{2}, - p \big )  S_\mathrm{os} (l)   \Gamma_{\pi} \big (k-\tfrac{p}{2}, p \big )  \Big  ] . 
\label{ff_func}   
\end{align}

We illustrate the relevant cut diagram of this FF in Fig.~\hyperlink{fig:diagram_b}{\ref{diagram}(b)} and note that Eqs.~\eqref{dly_relation2} and \eqref{ff_func} refer to a frame in which the produced hadron 
has no transverse momentum, $\boldsymbol{p}_T=0$, while the fragmenting quark does: $\boldsymbol{k}_T\neq 0$. To interpret this result as a distribution of a pion in the quark, 
one must boost to a frame where $\boldsymbol{k}_\perp=0$ and  $\boldsymbol{p}_\perp \neq 0$. This is discussed in detail in Refs.~\cite{Collins:1981uw,Barone:2001sp} and we 
can simply substitute $\boldsymbol{k}_T=-\boldsymbol{p}_\perp/z$. The elementary FF is normalized according to, 
\begin{eqnarray}
  \sum_{m}\int_0^1 \hat{d}_q^m (z) dz=1,\hspace{0.5cm}m=\pi^-,\pi^0,\pi^+ , 
\label{frag_sum_rule}
\end{eqnarray}
which ensures that the isospin and momentum sum rules for the full FF in Sec.~\ref{qjmodel} are satisfied~\cite{Ito:2009zc,Matevosyan:2010hh}.


\subsection{Quark-Jet Model\label{qjmodel}}

With the elementary FF~\eqref{ff_func} in hand, one must also consider the possibility that a quark can fragment via a cascading process of hadron emissions. 
The  quark-jet model of Field and Feynman~\cite{Field:1976ve} assumes that the meson observed in a semi-inclusive process is merely one amongst many in a  jet of hadrons. 
The probability of finding a pion with light-front momentum fraction $z$ in the jet is described by the full FF $D_q^\pi(z)$, also known as jet function.

The full FF can be conveniently written in terms of the isospin decomposition~\cite{Ito:2009zc},
\begin{equation}
    D_q^\pi(z) \equiv   \tfrac{1}{3}  \left [D_0(z)+\tau_q\tau_\pi D_1(z) \right ] ,
 \label{full_frag}
\end{equation}
where the functions $D_0$ and $D_1$ are given by Volterra integral equations:
\begin{eqnarray}
    \tfrac{2}{3} D_0^\pi(z)&=&\hat{d}_q^\pi(z)+\int_z^1 \frac{dy}{y} \ \hat{d}_u^\pi\qty(1-\frac{z}{y})D_0^\pi(y) \; , \\
    \tfrac{2}{3}D_1^\pi(z)&=&\hat{d}_q^\pi(z)-\frac{1}{3}\int_z^1 \frac{dy}{y} \ \hat{d}_u^\pi\qty(1-\frac{z}{y})D_1^\pi(y) \, .
    \hspace*{5mm}
\end{eqnarray}
In case of a $u$-quark fragmenting into a $\pi^+$, using the isospin conventions in Sec.~\ref{PDFsubsec}, Eq.~\eqref{full_frag} becomes $D_u^{\pi^+} = (D_0 + D_1)/3$.  In the limit $z \to 1$, 
the full fragmentation function reduces to the elementary one, $D_u^{\pi^+}(z) \overset{z \rightarrow 1}{\longrightarrow} \hat{d}_u^{\pi^+}(z)$, as all of the quark's momentum 
is transferred to the initial pion and no momentum is left for a jet. Finally, the moments of the full FFs are defined as:
\begin{eqnarray}
    \label{moments_FF}
    \expval{z}_{D_q^{\pi}}=\int_0^1 zD_q^\pi(z)dz\, .
\end{eqnarray}


\section{Theoretical Framework}
\label{sec:framework}

The quark gap equation in QCD is described by a DSE, a nonlinear integral equation that couples the two-point Green function of quark fermion fields, namely the fully dressed quark propagator,  with an infinite tower of higher-order Green functions~\cite{Bashir:2012fs},  
\begin{align} 
  S^{-1}(p) = & Z_2 \big(i\slashed{p}+m_\text{bm}\big)    \nonumber \\
          + & \  Z_1 g^2\! \int^\Lambda_k D_{\mu\nu}^{ab}(l)\frac{\lambda^a}{2}\gamma_\mu S(k)\Gamma_\nu^b(k,p) \, ,
 \label{quark_DSE}          
\end{align}
where $D^{ab}_{\mu\nu}(l)$ is the dressed gluon propagator with $l=k-p$,  $\Gamma_\nu^b(k,p)$ is the dressed quark-gluon vertex, $\lambda^a$ are SU(3) color matrices, $m_\textrm{bm}$ is the bare current-quark mass, and $Z_1(\mu,\Lambda)$ and  $Z_2(\mu,\Lambda)$ are the vertex and wave-function renormalization constants at the renormalization scale $\mu$, respectively. The regularization scale  $\Lambda \gg \mu$ can be taken to infinity. The covariant solution to this equation can generally be expressed as,
\begin{equation}
    S (p) = \frac{Z(p^2)}{i\slashed{p} + M(p^2)} , 
    \label{quark_prop_DS}      
\end{equation}
where $Z(p^2)$ is the wave renormalization function and $M(p^2)$ is the quark mass function. 

The bound-state interactions of a quark-antiquark pair in a pion are described by the Bethe-Salpeter equation (BSE) and in its homogeneous form 
is written as,
\begin{equation}
  \big[ \Gamma_\pi (k,p) \big]_{tu} =\!  \int_q  \big[K(k,q,p)\big]_{tu}^{rs} \big[S (q_\eta)\Gamma_\pi (q,p)S (q_{\bar{\eta}})\big]_{rs} \, ,
\label{BSE}
\end{equation}
where $K(k,q,p)$ is the fully amputated quark-antiquark scattering kernel and $r,...,u$ denote color, Dirac and flavor matrix indices. The quark and antiquark momenta are given by $q_{\eta} = q + \eta p$ and $q_{\bar \eta} = q - \eta p$, respectively, and the momentum-partitioning parameters satisfy $\eta+\bar{\eta}=1$. We stress that in a Poincar\'e covariant approach the mass spectrum and other static observables cannot depend on them due to translational invariance. 

The solution of the BSE~\eqref{BSE} for a bound state with quantum numbers  $J^{PC} = 0^{-+}$, relative momentum $k$ and total momentum $p$, is constrained by Lorentz covariance, parity and charge conjugation, and the most general BSA for a pseudoscalar meson reads~\cite{Llewellyn-Smith:1969bcu},
\begin{align} 
    \Gamma_\pi\qty(k,p) & = \gamma_5\big [ iE_\pi(k,p)+\slashed{p}\, F_\pi(k,p)   \nonumber \\
    & +  \slashed{k}\, k\cdot p\,  G_\pi(k,p)+\sigma_{\mu\nu} k^\mu p^\nu H_\pi(k,p)\big ] ,
\label{BSA_DS}
\end{align}
where $E_\pi$, $F_\pi$, $G_\pi$, and $H_\pi$ are Lorentz-invariant scalar amplitudes. The conjugate BSA is defined as $\bar{\Gamma}\pi\qty(k,p) = C\Gamma_\pi^T\qty(-k,p)C^T$, where $C$ is the charge-conjugation operator. Finally, we remind that the pion's weak decay constant is defined as:
\begin{equation}
\hspace*{-2mm}
   f_\pi p_\mu  =  \frac{Z_2 N_c}{\sqrt{2}} \!\! \int_q  \Tr_\text{\tiny D}\! \big [\gamma_5\gamma_\mu S (q_\eta)
                          \Gamma_\pi (q,p)  S (q_{\bar{\eta}}) \big ] .
    \label{pion_cte_DS}                      
\end{equation}


\subsection{Qin-Chang Model}
\label{sec:qc_model}

Working in the leading rainbow-ladder truncation, one simplifies $\Gamma_\mu^a\rightarrow Z_2\frac{\lambda^a}{2}\gamma_\mu$. This truncation is characterized by an abelianized Ward
identity which, as in QED, implies $Z_1=Z_2$~\cite{Bashir:2012fs}. With that, the kernel in the self-energy integral~\eqref{quark_DSE} can be written as, 
\begin{eqnarray}
  Z_1 g^2  D_{\mu\nu}(q)\Gamma_\nu^b(k,p) = Z_2^2\mathcal{G}(q^2)D_{\mu\nu}^{\text{\tiny free}}(q)\frac{\lambda^b}{2}\gamma_\nu \, ,
  \label{rainbow_truncation_DS}
\end{eqnarray}  
where the free-gluon propagator in Landau gauge is,
\begin{eqnarray}
   D_{\mu\nu}^{\text{\tiny free}}(q)=  \frac{1}{q^2}\bigg(\delta_{\mu\nu}-\frac{q_\mu q_\nu}{q^2}\bigg) \, .
   \label{free_gluon_QC}
\end{eqnarray}

The dressing function $\mathcal{G}(q^2)$ in Eq.~\eqref{rainbow_truncation_DS} is composed of two pieces,
\begin{equation}
    \frac{\mathcal{G} (q^2)}{q^2}  = \,  \mathcal{G}_\mathrm{IR} (q^2) +  4\pi \tilde\alpha_\mathrm{PT} (q^2)  \ ,
\label{IR+UV}    
\end{equation}
where a factor $1/q^2$ of the free-gluon propagator is included in the definition. The first term in Eq.~\eqref{IR+UV} dominates the infrared
momentum region,  $q^2 < \Lambda^2_\mathrm{QCD}$, and is suppressed at large momenta. The second term describes a monotonically decreasing continuation 
of the perturbative coupling that dominates at large momenta: $\mathcal{G}_{\mathrm{IR}} (q^2) \ll \tilde{\alpha}_{\mathrm{PT}} (q^2) \, \forall \, q^2 \gtrsim 2 \
\mathrm{GeV}^2$. We use the Qin-Chang\footnote{The name is a short reference to the authors Qin-Chang-Liu-Roberts-Wilson of Ref.~\cite{Qin:2011dd}.} model~\cite{Qin:2011dd},
which saturates in the infrared and can be associated with a dynamical gluon mass,
\begin{align}
    \mathcal{G}^\mathrm{IR} (q^2)  & =   \frac{8\pi^2}{\omega^4}  D\, e^{-q^2/\omega^2}    
     \label{G-IR}      \\
  4\pi \tilde\alpha_\mathrm{PT}(q^2) & =  \frac{8\pi^2  \gamma_m\, \mathcal{E}(q^2)}{ \ln \left [  \tau +
                       \left (1 + q^2/\Lambda^2_\textrm{\tiny QCD} \right )^{\!2} \right ] }  \ , 
\label{G-PT}                         
\end{align}
where $\gamma_m = 12/(33-2N_f)$ is the mass anomalous dimension, $N_f $ is the active flavor number, $\Lambda_\textrm{\tiny QCD}=0.234$~GeV, 
$\tau=e^2-1$, $\mathcal{E}(q^2)=[1-\exp(-q^2/4m^2_t)]/q^2$, $m_t=0.5$~GeV, $\omega= 0.5\,$GeV  and $\kappa=(\omega D)^{1/3} = 0.8\,$GeV.

The fully amputated scattering kernel compatible with rainbow truncation is given by the ladder truncation~\cite{Maris:1997hd}, 
\begin{equation}  
  [K(k,q,p)]_{tu}^{rs} = -Z_2^2\, \mathcal{G}(l^2)\, D_{\mu\nu}^{\text{\tiny free}}(l)\bigg(\frac{\lambda^a}{2}\gamma_\mu\bigg)_{tr} \bigg(\frac{\lambda^a}{2}\gamma_\nu\bigg)_{su}\!, 
  \label{kernel_QC}
\end{equation}
with the gluon momentum $l=k-q$. Inserting Eq.~\eqref{kernel_QC} in Eq.~\eqref{BSE}, we can rewrite the BSE as,
\begin{align}
    \Gamma_\pi(k,p)=-Z_2^2 C_F & \int_q \ \mathcal{G}(l^2)D_{\mu\nu}^{\text{\tiny free}}(l)\gamma_\mu S(q_\eta) \nonumber \\
    & \times \Gamma_\pi(q,p) S(q_{\bar{\eta}})  \gamma_\nu \, ,
\end{align}
where $C_F = (N_c^2 - 1)/2N_c$ is the Casimir Operator.


\subsection{Curci-Ferrari Model}
\label{sec:curci_ferrari}

The CF model~\cite{Curci:1976bt}, originally proposed as an alternative to the Higgs mechanism, can be understood as a consistent massive vector-boson extension of QCD. It has been revived in the past decade to explore nonperturbative effects and the infrared regime of QCD. In the original CF approach, a mass term~\cite{Curci:1976bt,Pelaez:2021tpq} is introduced at tree level in a gauge-fixed version of the Lagrangian, which guarantees its perturbative renormalizability and regulates the long-distance behavior.

We here consider the CF Lagrangian in Euclidean space and in Landau gauge with the presence of a Nakanishi-Lautrup field, $h_a$, that simplifies the writing of a 
modified BRST symmetry: 
\begin{eqnarray}   
  \mathcal{L}&=&\sum_{i=1}^{N_f}\bar{\Psi}_i(\slashed{D}+\mathcal{M}_\Lambda)\Psi_i+\frac{1}{4}F^a_{\mu\nu}F^a_{\mu\nu}  \nonumber    \\
   &+ & ih^a\partial_\mu A^a_\mu+\partial_\mu \bar{c}^a(D_\mu c)^2+\frac{1}{2}m_\Lambda^2(A_\mu^a)^2 \, .
\label{CF_lagrangian}   
\end{eqnarray}
In this Lagrangian, the usual covariant derivative is defined as $D_\mu = \partial_\mu -ig_\Lambda A_\mu^a$, 
$F_{\mu\nu}^a \equiv \partial_\mu A_\nu^a - \partial_\nu A_\mu^a + g_\Lambda f^{abc} A_\mu^b A_\nu^c$ is the field-strength tensor, $c^a$ and $\bar{c}^a$ are ghost and antighost fields and $\psi_i$ and $\bar{\psi}i$ are quark and antiquark fields of flavor $i$, respectively. The additional 
term, $\frac{1}{2} m^2_\Lambda (A_\mu^a)^2$, is motivated by dynamical chiral symmetry breaking and the emergence of a \emph{dynamical} gluon
 mass~\cite{Aguilar:2004sw,Aguilar:2008xm,Cucchieri:2009zt,Bogolubsky:2009dc,Boucaud:2008ky,Duarte:2016iko,Duarte:2017wte,Dudal:2018cli}.

A more sophisticated approach to the CF model combines an expansion in the inverse number of colors $N_c$ with an expansion of the strong coupling in the gauge sector,
while keeping a realistic running of the quark-gluon coupling and gluon mass at next-to-leading order. At leading order the well-known rainbow truncation emerges of this 
expansion with constant quark-gluon coupling and gluon mass. The systematic higher-order corrections in this expansion scheme, which implements renormalization group 
running at leading order, is referred to as \emph{rainbow-improved} loop expansion~\cite{Pelaez:2017bhh,Pelaez:2020ups,Pelaez:2022rwx}. In the DSE language this corresponds partially to the beyond 
\emph{rainbow-ladder} truncation that has received considerable attention over the past years~\cite{Rojas:2013tza,Williams:2015cvx,Fischer:2008wy,Albino:2018ncl,
Albino:2021rvj,El-Bennich:2022obe,Lessa:2022wqc,Oliveira:2018ukh,Oliveira:2020yac,Aguilar:2016lbe,Aguilar:2018epe,Aguilar:2024ciu,Gao:2021wun}.

The Feynman rules applied to the CF model lead to a leading-order gluon propagator in Landau gauge that includes the  mass term of the Lagrangian~\eqref{CF_lagrangian}, and the dressed gluon propagator is therefore, 
\begin{eqnarray}
   D_{\mu\nu}^{\text{\tiny CF}} (q)=  \frac{1}{q^2+m_\Lambda^2}\bigg(\delta_{\mu\nu}-\frac{q_\mu q_\nu}{q^2}\bigg) \, .
   \label{free_gluon_CF}
\end{eqnarray}
Adjusting the mass scale $m_\Lambda$ and the renormalized quark-gluon coupling in the  two-loop corrections of the quark self energy~\cite{Pelaez:2022rwx}, 
the wave function renormalization, $Z(p^2)$, and mass function,  $M(p^2)$, are momentum-dependent functions and the  dressed quark propagator is of the same form as in Eq.~\eqref{quark_prop_DS}.

The BSE for a pion in the CF model was solved in the chiral limit in Ref.~\cite{Pelaez:2022rwx} and reads for a given isospin, 
\begin{equation}
\hspace*{-1.5mm}
\gamma_\pi (k,p)  =  - C_F g^2\!\! \int_q D_{\mu\nu}^{\text{\tiny CF}}(l)\gamma_\mu S (q_\eta)  \gamma_\pi (q,p)S (q_{\bar{\eta}}) \gamma_\nu\, , 
\label{BSE_CF}	
\end{equation}
where $g$ is the renormalized strong coupling and again $C_F = (N_c^2 - 1)/2N_c \sim N_c/2$ for $N_c \gg 1$. Working in the chiral limit, the quark propagator~\eqref{quark_prop_DS} is expanded around  $p^2 = 0$ up to linear order in $p$ in Eq.~\eqref{BSE_CF}, which reduces the BSE to a set of one-dimensional coupled integral equations that can be solved numerically. As before, the most general BSA can be decomposed in terms of covariant amplitudes,
\begin{align}
	\gamma_\pi (k, p)&= -i \gamma_5 \left [  \gamma_P (k^2)+i\slashed{p}\,\gamma_A(k^2)+i\sigma_{\mu\nu}p_\mu k_\nu\gamma_T(k^2)  \right. \nonumber \\
	& + \, \left. 2 i\slashed{k}\, \frac{k\cdot p}{k^2} \big (  \gamma_A(k^2)-\gamma_B(k^2) \big )  \right  ] +\mathcal{O}(p^2)\, ,
\label{BSA_CF}
\end{align}
and $\gamma_P$, $\gamma_A$, $\gamma_B$ and $\gamma_T$ are real scalar amplitudes that depend solely on the relative momentum $k$ for $p^2 \to 0$, while the integral equation for $\gamma_P (k^2)$ decouples from the remaining, coupled ones. Numerical results for these scalar amplitudes are given in Fig. 5 and 6 of Ref.~\cite{Pelaez:2022rwx}.

The BSA in Eq.~\eqref{BSA_CF} can be related to the one in Eq.~\eqref{BSA_DS}, i.e. we write $E_\pi(k,P)$, $F_\pi(k,P)$, $G_\pi(k,P)$, and $H_\pi(k,P)$ in terms of 
$\gamma_P$, $\gamma_T$, $\gamma_A$ and $\gamma_B$: 
\begin{align}
	\label{bsaE_CF}
	E_\pi^{\text{\tiny CF}}(k^2) &\equiv-\frac{\sqrt{2}}{f_\pi}\gamma_P(k^2) \, , \\
	\label{bsaF_CF}
	F_\pi^{\text{\tiny CF}}(k^2) &\equiv \frac{\sqrt{2}}{f_\pi}\gamma_A(k^2) \, , \\
	\label{bsaG_CF}
	G_\pi^{\text{\tiny CF}}(k^2) &\equiv \frac{2\sqrt{2}}{f_\pi k^2}\qty[\gamma_A(k^2)-\gamma_B(k^2)]\, , \\
	\label{bsaH_CF}
	H_\pi^{\text{\tiny CF}}(k^2) &\equiv -\frac{\sqrt{2}}{f_\pi}\gamma_T(k^2)\, .
\end{align}


\section{Results}
\label{sec:numericals}

In order to compute the PDF~\eqref{pdf_func} and FF~\eqref{ff_func} with the quark propagators and the pion BSA from either approaches in Sec.~\ref{sec:qc_model}
or  \ref{sec:curci_ferrari}, we rewrite the integration measure in terms of light-cone variables~\cite{El-Bennich:2012mkr}, 
\begin{equation}
   \int d^4k = \frac{1}{2}\int dk^+dk^-d\boldsymbol{k}_T \, , 
\end{equation}
and integrate over $k^+$ and $k^-$ using both Dirac delta functions in Eq.~\eqref{pdf_func}. The second delta function yields,
\begin{align}
	\delta(l^2+M_E^2) & = 
	 \delta  \big ( \boldsymbol{k}_T^2 - k^+ k^-\!\! -m_\pi^2 + k^+p^-\! +k^-p^+\! + M_E^2 \big )  \nonumber \\
	 & = \,  \frac{1}{p^+(1-x)} \delta \left (k^- -f (\bm{k}_T^2,x) \right )  , 
\end{align}
where $f(\boldsymbol{k}_T^2,x)$ is defined as, 
\begin{equation}
	f(\boldsymbol{k}_T^2,x)  =  \frac{\boldsymbol{k}_T^2+M_E^2-(1-x)m_\pi^2}{p^+(1-x) }\equiv k^- ,
\end{equation}
and the remaining $\bm{k}_T$ integration can be performed numerically. The same procedure is used to integrate over $k^+$ and $k^-$ in the FF~\eqref{ff_func} where
we apply a boost to the frame where the pion has transverse momentum, $\boldsymbol{k}_T = -\boldsymbol{p}_\perp / z$; see Refs.~\cite{Ito:2009zc,daSilveira:2024ddq} for details.

Since the solutions of the quark propagators and the BSAs are obtained numerically in Euclidean space, a convenient algebraic parametrization of their scalar covariant amplitudes
is required to perform the integrals over the Dirac delta functions. The quark dressing functions $Z(k)$ and $M(k)$ can be parametrized with a Pad\'e  approximant~\cite{Falcao:2022gxt}, 
such as,
\begin{eqnarray}
M(k^2)&=&\frac{a_0+a_1k^2+a_2k^4}{1+b_1k^2+b_2k^4+b_3k^6} \, ,    \label{pade_mass}   \\
Z(k^2)&=&\frac{c_0+c_1k^2+c_2k^4}{1+d_1k^2+d_2k^4}\, . 
\label{pade_renorm}
\end{eqnarray}
The coefficients of both Eqs.~\eqref{pade_mass} and \eqref{pade_renorm} are listed in Table~\ref{tab:pade}. 
%
\begin{table}[htb]
    \caption{Parameters for the quark mass function~\eqref{pade_mass} and wave renormalization function~\eqref{pade_renorm}.}
    \renewcommand{\arraystretch}{1.2}
    \begin{tabularx}{\linewidth}{c *{6}{>{\centering\arraybackslash}X}}
        \hline
        \hline
                &  $a_0$ &  $a_1$  &   $a_2$  &   $b_1$  & $b_2$ & $b_3$  \\
        \hline
        
        CF    & $0.332$ & $4.949$ & $-0.284$ & $16.58$  & $55.74$ & $-4.812$    \\
        QC    & $0.500$ & $-0.076$ & $0.0126$ &  $1.324$  &  $0.636$   &  $0.328$   \\
        \hline
    \end{tabularx}\\[1pt]
    \begin{tabularx}{\linewidth}{c *{5}{>{\centering\arraybackslash}X}}
        \hline
                &  $c_0$ &  $c_1$  &   $c_2$  &   $d_1$  & $d_2$  \\
        \hline
        CF    & $0.985$ & $30.40$ & $222.2$ &  $223.5$   &  $22.78$    \\
        QC    & $0.717$ & $2.518$ & $1.413$ &  $3.469$  &  $1.170$   \\
        \hline
        \hline
    \end{tabularx}
    \label{tab:pade}
\end{table}

The BSA, on the other hand, can be parametrized with a  Nakanishi integral representation~\cite{Nakanishi:1965zza,Nakanishi:1965zz} of the form,
\begin{equation}
   \mathcal{F}_i(k,p)  =  \sum_{j=1}^N\int_{-1}^1  \frac{U_j\Lambda_j^{2n_j}  \rho_j(\alpha)}{(k^2+\alpha k\cdot p+\Lambda_j^2)^{n_j} } \, d\alpha  \, ,
\label{nakanishi}   
\end{equation}
in which we define the spectral density $\rho_j(\alpha)$  as~\cite{Serna:2024vpn},
\begin{equation}
    \rho_j (\alpha)    =  \frac{\Gamma  \qty(\frac{3}{2})}{\sqrt{\pi}\Gamma(1)}\qty[C_0^{1/2}(\alpha) +  \sigma_j C_2^{1/2}(\alpha)],
\label{rho_i} 
\end{equation}
where $C_n^{1/2}(\alpha)$ are Gegenbauer polynomials of order $1/2$. Since the BSA in the CF model~\eqref{BSA_CF} is only known in the chiral limit, its angular dependence 
vanishes and the $k\cdot p$ term in the denominator of Eq.~\eqref{nakanishi} is dispensable, while the integrals over the spectral functions $\rho_i(\alpha)$ are trivial. Therefore, 
the BSA of the CF model can be parametrized with a simpler expression,
\begin{equation}
   \mathcal{F}_i(k^2)  =   \sum_{j=1}^N \frac{U_j\Lambda_j^{2n_j}}{(k^2+\Lambda_j^2)^{n_j}} \, ,
\label{nakanishi_CF}
\end{equation}
and the parameters of this CF-BSA are found in Tab.~\ref{tab:BSA_CF}.

On the other hand, the numerical solution of the BSA in Eq.~\eqref{BSA_DS} is obtained in the near-chiral limit, $m_\pi \approx 0.3$\;MeV, and it turns
out that even for such a small meson mass the angular dependence is to a certain extent relevant. We therefore expand the expression in Eq.~\eqref{nakanishi}
with $\Lambda_i = \Lambda$ up to order $z_k^2$, where $z_k = k\cdot p/|k||p|$, which yields, 
\begin{align}
  \mathcal{F}_i(k^2,z_k)  = & \sum_{j=1}^N\,  U_j  \Lambda^{2n_j}  \bigg [ \frac{ 1 }{ (k^2+\Lambda^2)^{n_j} } \nonumber \\
   & \  -\frac{k^2m_\pi^2 z_k^2 \ n_j(n_j+1)(5+2\sigma_j)}{ 30 (k^2+\Lambda^2)^{n_j+2}} \bigg ] \, ,
 \label{expandNakanishi}  
\end{align}
and use the Chebyshev expansion of the scalar amplitudes $E_\pi$, $F_\pi$, $G_\pi$ and $H_\pi$, 
\begin{equation}
   \mathcal{F}_i(k^2,k\cdot p)  = \sum_{m=0}^\infty\mathcal{F}_{im}(k^2)U_m(z_k) \, ,
\end{equation} 
keeping terms up to $z_k^2$ of the Chebyshev polynomials $U_m(z_k)$ of the second kind. Fitting the parameters $U_j, n_j$ and $\sigma_j$ for $N=4$ in Eq.~\eqref{expandNakanishi}, we obtain the values listed in Tab.~\ref{tab:BSA_QC}.
%
\begin{table*}[htb]
    \caption{Parameters of the BSA in the CF model.}
    \renewcommand{\arraystretch}{1.2}
    \begin{tabularx}{\linewidth}{c *{10}{>{\centering\arraybackslash}X}}
        \hline
        \hline
        $\mathcal{F}_i$ & $\Lambda_1$ & $\Lambda_2$ & $\Lambda_3$ &  $U_1$  &   $U_2$  &   $U_3$  &  $n_1$ & $n_2$ & $n_3$ \\
        \hline
        $E_\pi$         &  $-67.08$  & $-1.134$ & $3.15$ & $-0.199$ & $-3.943$ &  $-1.374$   &   $4$  &  $5$  & $6$ \\
        $F_\pi$         &  $-1.274$  & $0.983$ & $0.983$ &  $0.336$ & $2.014$ & $-2.397$ &  $2$  &  $6$  & $7$ \\
        $G_\pi$         &  $-0.362$  & $0.438$ & $0.282$ & $176.9$ & $-147.6$ & $-46.85$  &  $3$  &  $4$  & $5$ \\
        $H_\pi$         &  $0.747$  & $3.030$ & $-0.349$ & $-3.182$ & $-0.128$ & $3.343$ & $4$  &  $5$  & $6$ \\
        \hline
        \hline
    \end{tabularx}\\[0pt]    
    \label{tab:BSA_CF}
    \caption{Parameters of the BSA in the QC model.}
    \renewcommand{\arraystretch}{1.2}
    \begin{tabularx}{\linewidth}{c *{14}{>{\centering\arraybackslash}X}}
        \hline
        \hline
        $\mathcal{F}_i$ & $\Lambda$ &  $U_1$  &   $U_2$  &   $U_3$  &  $U_4$ & $\sigma_1$ & $\sigma_2$ & $\sigma_3$ &  $\sigma_4$ & $n_1$ & $n_2$ & $n_3$ & $n_4$ \\
        \hline
        $E_\pi$         &  $1.267$  & $2.571$ & $-1.572$ &  -- & --   &  $2.136$   &  $1.889$   &  --   & -- & $4$  &  $5$  & -- & -- \\
        $F_\pi$         &  $1.247$  & $0.007$ & $2.660$ & $-2.454$ &  $0.309$ &  $-76.88$  &  $-0.367$  &  $-0.020$  &  $1.024$ & $2$  &  $6$  & $7$ &  $10$ \\
        $G_\pi$         &  $0.971$  & $-0.874$ & $3.458$ &  $-2.096$   &  -- & $-3.289$  &  $-1.714$  &  $-1.741$    & -- &  $3$  &  $4$  & $5$ &  -- \\
        $H_\pi$         &  $1.156$  & $-0.207$ & $1.606$ & $-1.125$ & -- & $-39.54$  &  $-8.158$  &  $-6.261$  &  -- & $4$  &  $6$  & $7$ & --  \\
        \hline
        \hline
    \end{tabularx}
    \label{tab:BSA_QC}
\end{table*}

\begin{figure}[t!]
	\centering
	\includegraphics[scale=0.55,angle=0]{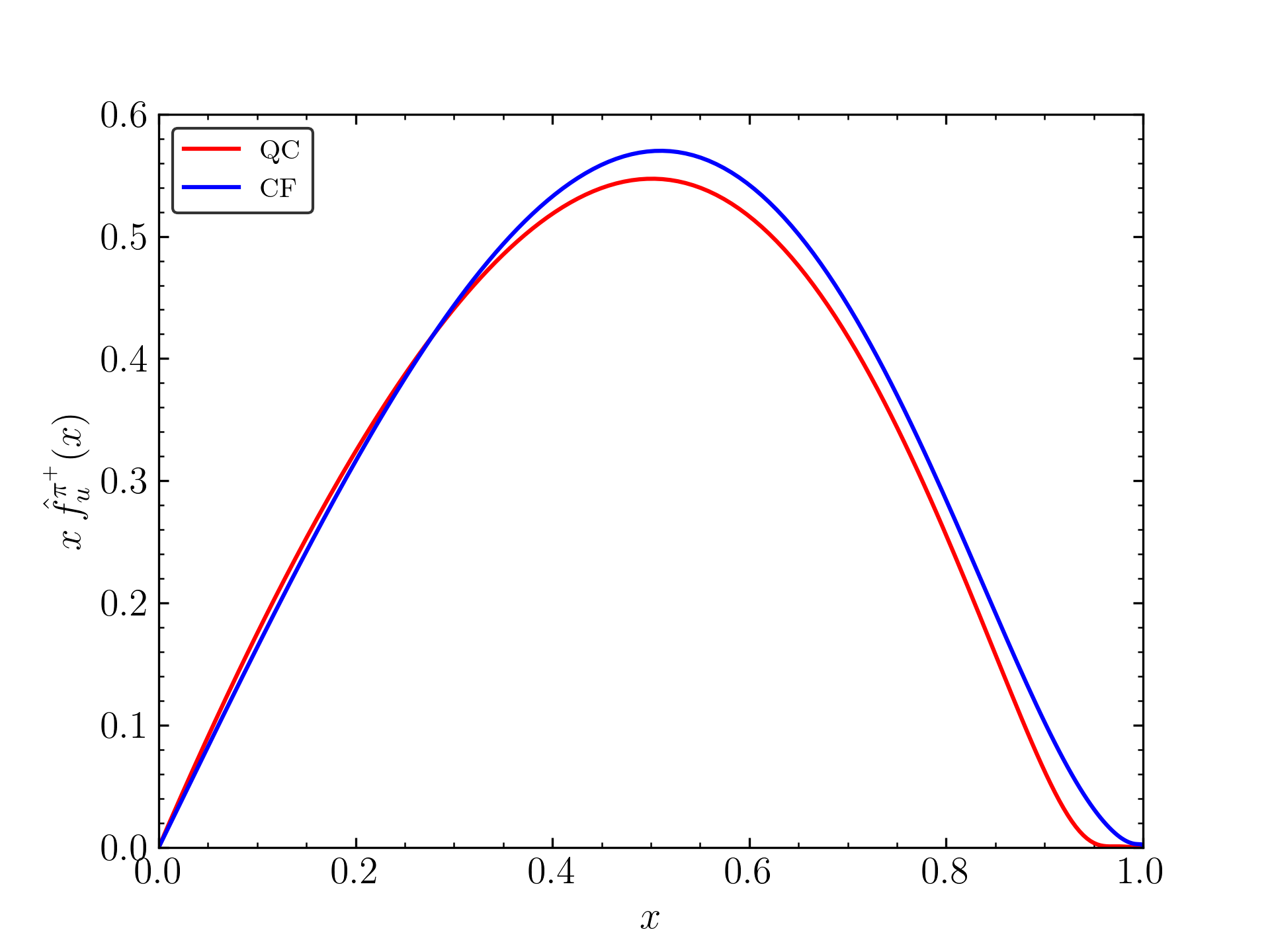}
	\caption{Normalized PDF computed with Eq.~\eqref{pdf_func}, the QC and CF frameworks.}
	\label{fig:pdf}
	\includegraphics[scale=0.55,angle=0]{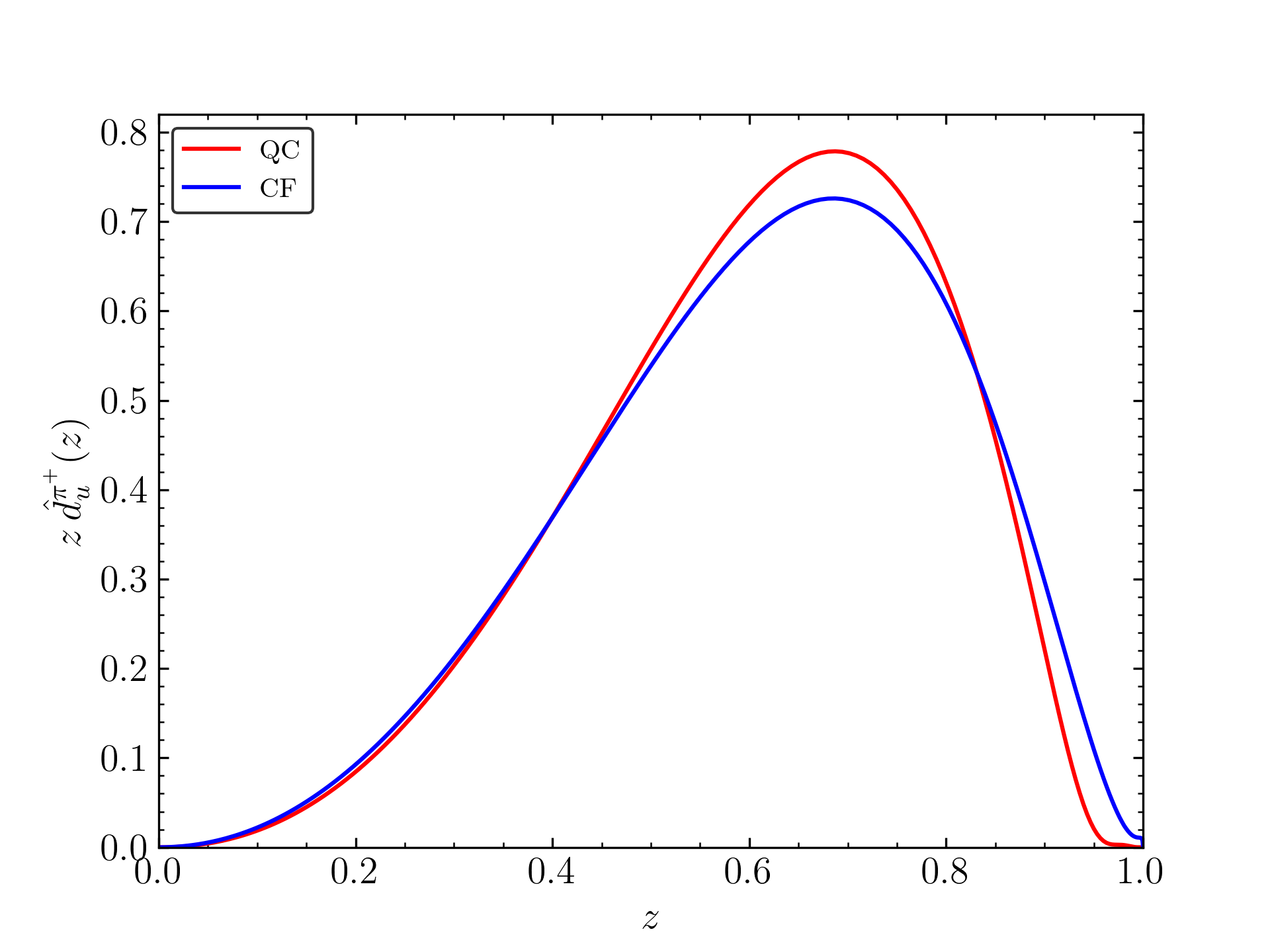}
	\caption{Normalized FF computed with Eq.~\eqref{ff_func}, the QC and CF frameworks.}
	\label{fig:frag}    
\end{figure}
\begin{figure*}[t!]
	\centering
	\includegraphics[scale=0.58,angle=0]{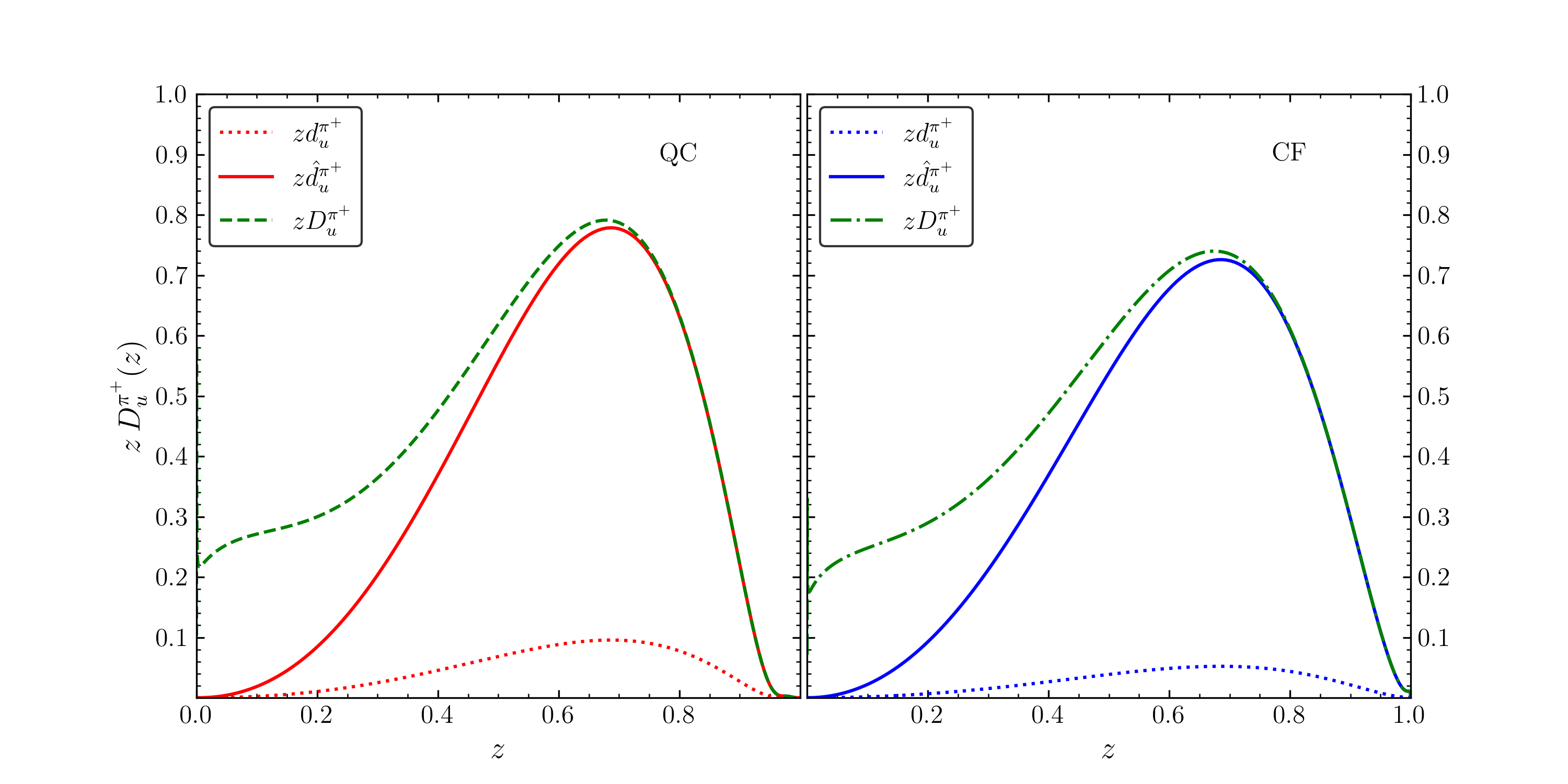}
		\vspace*{-5mm}
	\caption{Comparison of the (un)normalized elementary and full fragmentation functions, Eqs.~\eqref{ff_func} (dotted line), \eqref{frag_sum_rule} (solid line) 
	and  \eqref{full_frag} (dashed line). \emph{Left-hand side\/}: predictions of the QC model~\ref{sec:qc_model}; \emph{Right-hand side\/}: predictions of the 
	CF model~\ref{sec:curci_ferrari}. }
	\label{fig:full_frag}
\end{figure*}

We present our results obtained with the theoretical frameworks detailed in Sec~\ref{sec:framework} for the PDF and FF in Figs.~\ref{fig:pdf} and \ref{fig:frag}, respectively, from which we infer that the predictions of the semi-perturbative CF model and of the QC model are qualitatively and even quantitatively comparable. The only notable differences 
occur for $x \gtrsim 0.5$, where the distributions peak at very similar momentum fractions but the PDF and FF of the CF model fall off more slowly. In Fig.~\ref{fig:full_frag} 
the full FF~\eqref{full_frag} is compared to the normalized and unnormalized elementary FF,  $\hat d_u^\pi(z)$ and $d_u^\pi(z)$. Clearly, for values smaller than 
$z\approx 0.75$ the full fragmentation is similarly enhanced in both calculations and, as expected, one recovers the normalized elementary FF  in the limit of $z\rightarrow 1$.

We obtain the moments of the full FF using Eq.~\eqref{moments_FF}. For both approaches, we find
\begin{align}
    \label{moments_QC}
    \expval{z}^{\text{\tiny QC}}_{D_0^{\pi}} &=0.876\, , \hspace{0.15cm}
    \expval{z}^{\text{\tiny QC}}_{D_1^{\pi}}  =0.482\, , \hspace{0.15cm}
    \expval{z}^{\text{\tiny QC}}_{D_u^{\pi^+}}  =0.453\, ,  \\
    \label{moments_CF}
    \expval{z}^{\text{\tiny CF}}_{D_0^{\pi}} &  =0.848\, , \hspace{0.15cm}
    \expval{z}^{\text{\tiny CF}}_{D_1^{\pi}}   =0.480\, , \hspace{0.15cm}
    \expval{z}^{\text{\tiny CF}}_{D_u^{\pi^+}}  =0.443\, .
\end{align}

Consistent with what we observe in Fig.~\ref{fig:full_frag}, the CF model yields a full FF that is slightly suppressed compared with that in he QC model. This is reflected in the lower $\expval{z}_{D_u^{\pi^+}}$ value in Eq~\eqref{moments_CF}.

The renormalization scale of our calculations is determined such that the momentum fraction carried by the valence quarks agrees with the first moment of a  $\pi N$ Drell-Yan analysis, $2 \langle x \rangle_v = 0.47(2)$~\cite{Sutton:1991ay,Gluck:1999xe} at a scale $Q = 2\;$GeV. We do so by matching the moments using next-to-leading order Dokshitzer-Gribov-Lipatov-Altarelli-Parisi evolution~\cite{Botje:2010ay} including strange- and charm-quark mass thresholds from the experimental scale $Q$ to our model scale $Q_0$. We find for the QC model calculation $Q_0 = 0.77\;$GeV, and for the CF-model  $Q_0 = 0.73\;$GeV.


\section{Conclusions}
\label{sec:conclusions}

We computed the PDF and FF of the pion using two different gluon-dressing models within the same framework of gap and bound-state equations: the QC and the CF gluon models. We restricted ourselves to the massless pion, as the BSA in the CF approach is known in the chiral limit only~\cite{Pelaez:2022rwx}.  Our main goal was to verify whether a calculation with the massive CF propagator reproduces the distribution and fragmentation functions obtained with the QC model. It turns out that both calculations are not only in good qualitative agreement, but also display remarkable quantitative consistency. Given that the pion PDF in Ref.~\cite{Serna:2024vpn} compares favorably with experiment, these findings support the validity of the CF model as an effective, dynamical chiral symmetry breaking approach that can be used to study the internal momentum distribution of mesons and possibly baryons in the future.

The present study can be extended beyond the chiral limit once the BSA of the physical pion in the CF model is known. This would allow for a direct comparison with a large body of work within the DSE approach.  Moreover, the NJL model has been used to obtain the elementary FFs of the pion, kaon, and $\rho$ mesons and to couple their full FFs~\cite{Matevosyan:2010hh,Matevosyan:2011ey,Matevosyan:2011vj}. A similar study in the CF framework requires the computation of the corresponding BSAs of these bound states, from which we can derive transverse-momentum-dependent distribution and fragmentation functions and further test the robustness of the CF model.


\acknowledgments 

This work was supported by Coordena\c{c}\~ao de Aperfei\c{c}oamento de Pessoal de N\'{\i}vel Superior (CAPES), grants nos.~88887.935044/2024-00 (G.B.B.) and 88887.086709/2024-00 (R.C.S). Additional partial support was provided by Funda\c{c}\~ao de Amparo \` a Pesquisa do Estado de S\~ao Paulo (FAPESP), grants nos.~2023/00195-8 (B.~B) and 2018/25225-9 (G.K.), and Conselho Nacional de Desenvolvimento Cient\'{\i}fico e Tecnol\'ogico, grants nos.~409032/2023-9 (B.B) and 309262/2019-4 (G.K.). F.~E.~S. acknowledges support by the ICTP South American Institute for Fundamental Research during his visit. 


\end{document}